\documentclass[a4paper,aps,11pt,preprint,showpacs]{revtex4}
\usepackage{graphicx,amsmath,revsymb,bm,amssymb,amsfonts,amsthm}
\begin{document}

\title{Photoionization of helium-like ions in asymptotic nonrelativistic
region}
\author{A.I.~Mikhailov$^{\,\mathrm{a,b}}$, A.V.~Nefiodov$^{\,\mathrm{a,c}}$,
G.~Plunien$^{\,\mathrm{c}}$}
\affiliation{$^{\mathrm{a}}$Petersburg Nuclear Physics Institute,
188300 Gatchina, St.~Petersburg, Russia \\
$^{\mathrm{b}}$Max-Planck-Institut f\"ur Physik komplexer Systeme,
N\"othnitzer Stra{\ss}e 38, D-01187 Dresden, Germany \\
$^{\mathrm{c}}$Institut f\"ur Theoretische Physik, Technische
Universit\"at Dresden, Mommsenstra{\ss}e 13, D-01062  Dresden,
Germany }

\date{Received \today}
\widetext
\begin{abstract}
The cross section for single K-shell ionization by a high-energy
photon is evaluated in the next-to-leading order of the
nonrelativistic perturbation theory with respect to the
electron-electron interaction. The screening  corrections are of
particular importance for light helium-like ions. Even in the case
of neutral He atom, our analytical predictions turn out to be in
good agreement with the numerical calculations performed with the
use of the sophisticated wave functions. The universal high-energy
behavior is studied for the ratio of double-to-single
photoionization cross sections. We also discuss the fast convergence
of the perturbation theory over the reversed nuclear charge number
$1/Z$.
\end{abstract}
\maketitle

The single and double photoeffects on the helium isoelectronic
sequence represent the simplest fundamental processes, which are
being intensively investigated during last decades
\cite{1,2,3,4,5,6}. The accurate treatment of electron correlations
is still one of the main theoretical problems. Due to recent
developments of novel synchrotron radiation sources, the study of
the ionization of inner-shell electrons by high-energy photons is of
particular interest.

For the theoretical description of the ionization processes on light
atomic systems, it is usual to employ sophisticated methods with
highly correlated wave functions. This allows one to take into
account electron correlation effects beyond the independent-particle
approximation. However, all the methods suffer from the gauge
dependence. The latter can serve as a level of accuracy for the
theoretical predictions. In addition, the final results for cross
sections of the ionization processes are presented in a numerical
form, which is not always easy to analyze.

In the case of heavy multicharged ions, on the contrary, the usual
starting point is the approximation of non-interacting electrons,
which are described by the Coulomb wave functions for the discrete
and continuous spectra. The electron-electron interaction is treated
within the framework of perturbation theory, which is also referred
to as the expansion with respect to the parameter $1/Z$. The latter
represents the ratio of the strength of the electron-electron
interaction to the electron-nucleus one. To leading orders,
perturbation theory allows one to derive analytical results.
Accounting for higher-order correlation corrections improves the
accuracy of the analytical predictions in the domain of lower values
of the nuclear charge number $Z$. The results obtained within the
framework of perturbation theory for the binding energies and for
the cross sections are gauge independent.

In this Letter, we evaluate the next-to-leading-order
correlation correction to the cross section for single K-shell
ionization at asymptotic photon energies $\omega$ characterized by
$I \ll \omega \ll m$, where $I=\eta^2/(2m)$ is the Coulomb potential
for single ionization with $\eta = m\alpha Z$ being the average
momentum of a K-shell electron, $m$ is the electron mass, and
$\alpha$ is the fine-structure constant ($\hbar=1$, $c=1$). The
ejected electrons are considered as being nonrelativistic.
Accordingly, the Coulomb parameter is supposed to be sufficiently
small, that is, $\alpha Z \ll 1$.

Neglecting terms of order $(\alpha Z)^2$, the operator describing
the electron-photon interaction reads \cite{7}
\begin{equation}
\hat{V}_{\gamma} =  N_{\gamma} e^{i(\bm{k} \cdot \bm{r})}
\left(\bm{\mathrm e} \cdot \hat{\bm{p}} \right) \, , \qquad
N_{\gamma}= \frac{1}{m}\frac{\sqrt{4\pi \alpha}}{\sqrt{2\omega}} \,
.   \label{eq1}                                            
\end{equation}
Here $\hat{\bm{p}}$ is the momentum operator of an electron, which,
in the coordinate representation, is cast into the gradient form
$\hat{\bm{p}}= - i \bm{\nabla}$. An incoming photon is characterized
by the momentum $\bm{k}$, the energy $\omega = |\bm{k}| = k $, and
the polarization vector $\bm{\mathrm e}$. We employ the Coulomb
gauge, in which $(\bm{\mathrm e}\cdot \bm{k}) = 0$ and $(\bm{\mathrm
e}^*\cdot \bm{\mathrm e}) = 1$. In general, the nonrelativistic
interaction between an electron and a photon includes also
spin-dependent terms. However, in the case of single and double
photoeffects the corresponding contributions to the cross sections
are strongly suppressed \cite{8,9,10} and, therefore, can be
neglected.

In the nonrelativistic approximation, spatial and spin parts of
two-electron wave functions factorize. Moreover, the operator
\eqref{eq1} does not involve the spin matrices. As a result, the
spin functions can be omitted throughout this consideration, while
the symmetry of the coordinate wave functions $\Psi(\bm{r}_1,
\bm{r}_2)$ is preserved in the ionization process. The total
amplitude of the single photoeffect on helium-like ion is given by
\begin{equation}
{\cal A}=2 \,\langle \Psi_f | \hat{V}_{\gamma}|\Psi_i \rangle \,  ,
\label{eq2}                                                
\end{equation}
where the factor $2$ accounts for the one-particle character of the
operator \eqref{eq1}. That is, it is sufficient to consider the
interaction of an incoming photon with a single atomic electron
only.

In first-order perturbation theory with respect to the
electron-electron interaction, the wave functions $\Psi_{i,f}$ of
the initial and final states are represented as follows $\Psi_{i,f}
\simeq \Psi^{(0)}_{i,f} + \Psi^{(1)}_{i,f}$.  Accordingly, the
amplitude of the process is just ${\cal A} \simeq {\cal A}^{(0)} +
{\cal A}^{(1)}$. Neglecting the electron-electron interaction, we
shall employ the single-particle approximation in the external
Coulomb field of the nucleus (Furry picture):
\begin{eqnarray}
\Psi_i^{(0)}(\bm{r}_1,\bm{r}_2)&=&
\psi_{1s}(\bm{r}_1)\psi_{1s}(\bm{r}_2)  \, ,
\label{eq3}                                      \\       
\Psi_f^{(0)}(\bm{r}_1,\bm{r}_2)&=& \frac{1}{\sqrt{2}} \left[
\psi_{\bm{p}}(\bm{r}_1)\psi_{1s}(\bm{r}_2) +
\psi_{\bm{p}}(\bm{r}_2)\psi_{1s}(\bm{r}_1) \right] \,  .
\label{eq4}                                               
\end{eqnarray}
Here $\bm{p}$ is the momentum of escaping electron at infinity. The
explicit expressions for the first-order corrections
$\Psi^{(1)}_{i,f}$ to the wave functions can be found in works
\cite{10,11}.

\begin{figure}[h]
\centerline{\includegraphics[scale=0.6]{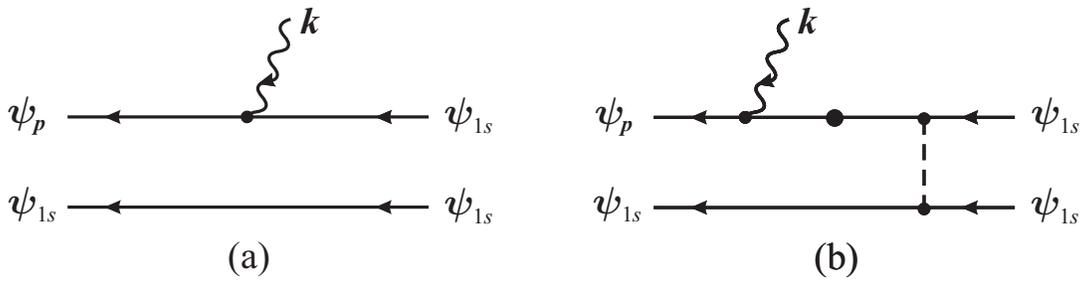}}
\caption{\label{fig1} Feynman diagrams for the single ionization of
K-shell electrons by a single photon. Solid lines denote electrons
in the Coulomb field of the nucleus, the dashed line denotes the
electron-electron Coulomb interaction, and the wavy line denotes an
incident photon. The line with a heavy dot corresponds to the
reduced Coulomb Green's function. Diagram (a) does not account for
the electron-electron interaction, while diagram (b) accounts for it
in the initial state.}
\end{figure}

In zeroth approximation, the amplitude \eqref{eq2} looks as follows
\begin{equation}
{\cal A}^{(0)}= 2 \,\langle \Psi_f^{(0)} |
\hat{V}_{\gamma}|\Psi_i^{(0)} \rangle = \sqrt{2} \,\langle
\psi_{\bm{p}} |\hat{V}_{\gamma}| \psi_{1s} \rangle \, .
\label{eq5}                                                
\end{equation}
The matrix element \eqref{eq5} can be represented by the Feynman
graph depicted in Fig.~\ref{fig1}(a). Apart from the common factor,
the expression \eqref{eq5} coincides with the amplitude for single
photoeffect on hydrogen-like ion in the ground state.

In the asymptotic region, the energy and momentum conservation laws
for the single K-shell photoionization keep the same form for both
H- and He-like ions:
\begin{eqnarray}
E_p &=& E_{1s} + \omega \simeq \omega \, ,
\label{eq6}                                      \\       
\bm{p}&=&  \bm{q} + \bm{k} \,  .
\label{eq7}                                               
\end{eqnarray}
Here $E_{1s} =- I$ is the Coulomb energy of the K-shell electron and
$-\bm{q}$ is the recoil momentum transferred to the nucleus.

Let us evaluate the amplitude \eqref{eq5} using the momentum
representation. Within the Born approximation, the wave function of
the ejected high-energy electron is described by a plane wave, that
is,
\begin{equation}
\langle \bm{f} |  \psi_{\bm{p}}\rangle  \simeq \langle \bm{f} |
\bm{p} \rangle = (2\pi)^3 \delta(\bm{f} -\bm{p}) \, .
\label{eq8}                                                
\end{equation}
Here the standard normalization on $\delta$ function in the momenta
is employed. Then the matrix element \eqref{eq5} yields
\begin{equation}
{\cal A}^{(0)}= \sqrt{2} \, N_{\gamma} \langle \bm{q}| \psi_{1s}
\rangle  \left(\bm{\mathrm e} \cdot \bm{p} \right)  \, .
\label{eq9}                                                
\end{equation}
If $q \gg \eta$ holds, the Coulomb wave function of the K-shell
electron reads
\begin{equation}
\langle \bm{q}| \psi_{1s}  \rangle  \simeq  N_{1s}  \frac{8 \pi
\eta}{q^4} \,  ,
\label{eq10}                                                
\end{equation}
where $N^2_{1s}=\eta^3/\pi$. Since in the nonrelativistic domain the
momentum $\bm{k}$ of a photon is negligibly small with respect to
the electron momentum $\bm{p}$, relation \eqref{eq7} can be written
as
\begin{equation}
\bm{p} \simeq  \bm{q} \,  .
\label{eq11}                                                
\end{equation}
The latter is equivalent to the use of the dipole approximation.
Then one can set $q^2 = p^2 = 2m \omega$ into Eq.~\eqref{eq10}. The
amplitude \eqref{eq9} can be further simplified. The corresponding
expression for the total cross section is well known \cite{7}
\begin{equation}
\sigma^+_{0}  = \frac{2^8 \pi \alpha}{3 m \omega}
\left(\frac{I}{\omega}\right)^{5/2} \,  , \qquad (I \ll \omega \ll
m) \, .
\label{eq12}                                             
\end{equation}

If the photon energy is not too high, as the wave function of the
final state one needs to utilize the one-electron Coulomb wave
function of the continuous spectrum. Accordingly, within the dipole
approximation the amplitude \eqref{eq5} leads to the following
expression for the total cross section,
\begin{equation}
\sigma^+_{\mathrm{C}} = \alpha  a_0^2 \,\frac{2^{10}\pi^2}{3Z^{ 2} }
\frac{\exp(-4\xi \cot^{-1}\xi )}{\varepsilon^4_\gamma [1 - \exp
(-2\pi \xi)]}   \,  , \qquad (\omega \ll m) \, ,
\label{eq13}                                                
\end{equation}
which is in fact valid in the whole nonrelativistic domain \cite{7}.
Here $\varepsilon_{\gamma} = \omega/I$ denotes the dimensionless
energy for the incident photon, $a_0 = 1/(m \alpha)$ is the Bohr
radius, and $\xi = 1/\sqrt{\varepsilon_{\gamma} -1}$. The
dimensionless parameter $\xi^{-1} = p/\eta$ has the meaning of the
momentum $p$ of the ejected electron, which is calibrated in units
of the characteristic momentum $\eta$. Formula \eqref{eq12} provides
just the leading term in the expansion of Eq.~\eqref{eq13} with
respect to the parameter $\xi \ll 1$. Since the expression
\eqref{eq13} involves the combination $\pi \xi$, the convergence of
the $\xi$ expansion is slow. Note also that both cross sections
\eqref{eq12} and \eqref{eq13} are twice as large as that for the
single photoeffect on a hydrogen-like ion in the ground state. This
is the consequence of the approximation of non-interacting electrons
employed in the derivation. Neither $\sigma^+_{0}$ nor
$\sigma^+_{\mathrm{C}}$ account for electron correlation effects.

Now we shall consider in more details the evaluation of the
next-to-leading-order correction ${\cal A}^{(1)}$ to the amplitude
of single K-shell photoeffect on helium-like ion. In the high-energy
nonrelativistic limit, the dominant contribution to the amplitude of
the process arises only from the Feynman diagram depicted in
Fig.~\ref{fig1}(b), providing the Coulomb gauge is employed. This
graph accounts for the interaction between the electrons in the
initial state. All other diagrams, namely, the one, which accounts
for the electron-electron interaction in the final state, together
with both exchange diagrams, turn out to be suppressed by the factor
of about $I/\omega$ and, therefore, can be neglected. Accordingly,
we can write
\begin{equation}
{\mathcal A}^{(1)} = \sqrt{2} \, \langle \psi_{\bm{p}}  \psi_{1s} |
\hat{V}_{\gamma} G_{\mathrm{R}}(E_{1s}) V_{12} | \psi_{1s} \psi_{1s}
\rangle \, .
\label{eq14}                                            
\end{equation}
Here the operator $V_{12}$ describes the Coulomb electron-electron
interaction. In the coordinate representation, it reads
\begin{equation}
V_{12} = \frac{\alpha}{|\bm{r}_1 - \bm{r}_2|}  \,  .
\label{eq15}                                            
\end{equation}
The reduced Green's function $G_{\mathrm{R}}(E_{1s})$ corresponding
to the energy $E_{1s}$ of the K-shell electron is related to the
usual nonrelativistic Coulomb Green's function $G_{\mathrm{C}}(E)$
as follows
\begin{equation}
G_{\mathrm{R}}(E_{1s}) = \lim_{E \to E_{1s}} \left\{
G_{\mathrm{C}}(E) - \frac{ |\psi_{1s} \rangle  \langle \psi_{1s}|
}{E - E_{1s}} \right\}   \,  .
\label{eq16}                                            
\end{equation}

Within the Born approximation \eqref{eq8}, the amplitude
\eqref{eq14} yields
\begin{eqnarray}
{\mathcal A}^{(1)} &=&  4 \pi \alpha \sqrt{2}\, N_{\gamma}
\left(\bm{\mathrm e} \cdot \bm{p} \right) \int \langle \bm{q} |
G_{\mathrm{R}}(E_{1s}) |\bm{f}_1 \rangle \frac{1}{\bm{f}^{ 2}}
\langle \bm{f}_1 + \bm{f} |\psi_{1s}
\rangle   \times \nonumber\\
&&\times \langle \psi_{1s} | \bm{f}_2\rangle \langle \bm{f}_2 -
\bm{f} |\psi_{1s} \rangle \frac{d\bm{f}_1}{(2\pi)^3}
\frac{d\bm{f}_2}{(2\pi)^3}  \frac{d\bm{f}}{(2\pi)^3} \,   ,
\label{eq17}                                            
\end{eqnarray}
where $\bm{q}= \bm{p} - \bm{k}$. Integrating over the intermediate
momenta in Eq.~\eqref{eq17}, one receives
\begin{eqnarray}
{\mathcal A}^{(1)} &=&  4 \pi \alpha  \sqrt{2} \, N_{\gamma}
N_{1s}^3 \left(\bm{\mathrm e} \cdot \bm{p} \right)
\times\nonumber\\
&&\times \left( - \frac{\partial}{\partial \mu} \right)
\frac{1}{\mu^2} \langle \bm{q} |  G_{\mathrm{R}}(E_{1s}) \left(
V_{i\eta} - V_{i(\eta +\mu)} \right)| 0 \rangle_{\left| \mu = 2\eta
\right. } \,   ,  \label{eq18} \\
\langle \bm{f}'| V_{i\lambda} | \bm{f}  \rangle &=& \frac{4\pi}{(
\bm{f}'- \bm{f} )^2 +\lambda^2} \,  . \nonumber
\end{eqnarray}
After taking the derivative with respect to $\mu$, one should set
$\mu = 2\eta$, where $\eta = m \alpha Z$.

In Eq.~\eqref{eq18}, we shall evaluate first the matrix element with
the Coulomb Green's function. Since $q \gg \eta$, one can employ the
integral representation \cite{12}
\begin{eqnarray}
\langle \bm{q} |G_{\mathrm{C}}(E) V_{i\lambda}| 0 \rangle &\simeq &
4 m \eta\frac{i p_1}{q^4} \int_1^{\infty} \left( \frac{y +1}{y -1}
\right)^{i\zeta} \langle 0 | V_{p_1 y + i\lambda} | 0  \rangle \, dy
=  \label{eq19} \\
&=& 2^5 \pi m \eta \frac{ip_1}{q^4} \int_0^1 \frac{ t^{-i\zeta } \,
dt}{[\lambda(1-t) - ip_1(1+t)]^2} \,  , \label{eq20}
\end{eqnarray}
where $p_1=\sqrt{2 m E}$ and $\zeta= \eta/p_1$. In order to isolate
the pole contribution together with finite terms at $E \to E_{1s}$
in Eq.~\eqref{eq20}, we set $E=E_{1s} - \delta$. The shift $\delta$
is supposed to be small and positive. Expanding the intermediate
momentum $p_1$ into a series over the parameter $\delta$, one
receives
\begin{eqnarray}
p_1 &\simeq& i\eta(1 + \varepsilon - \varepsilon^2/2) \,   ,
\label{eq21} \\
1 - i\zeta &\simeq& \varepsilon (1 - 3\varepsilon/2) \, ,
\label{eq22}
\end{eqnarray}
where $\varepsilon=\delta/(2 I)$. Performing the integration in
Eq.~\eqref{eq20} by parts and using the expansions \eqref{eq21} and
\eqref{eq22}, we find that
\begin{eqnarray}
\frac{ip_1}{\eta} \int_0^1 \frac{ t^{-i\zeta}\, dt}{[\lambda(1-t) -
ip_1(1+t)]^2} &\simeq& - \frac{1}{(\lambda +\eta)^2} \left(
\frac{1}{ \varepsilon} + \frac32 + \frac{\lambda -\eta}{\lambda
+\eta} \right)  +  \label{eq23} \\
&& + 2(\lambda -\eta) \int_0^1 \frac{\ln t \,  dt }{[\lambda +\eta -
(\lambda -\eta)t ]^3}   \,  .  \label{eq24}
\end{eqnarray}
In the integral \eqref{eq24} we have set $\delta =0$, which is
equivalent to the substitution $ip_1 \to - \eta$. As a result, the
matrix element \eqref{eq19} can be cast into the form
\begin{eqnarray}
\langle \bm{q} |G_{\mathrm{C}}(E_{1s} - \delta) V_{i\lambda}| 0
\rangle &=& 2^5 \pi m \frac{\eta^2}{q^4}\left\{- \frac{1}{(\lambda
+\eta)^2} \left( \frac{1}{ \varepsilon} + \frac32 + \frac{\lambda
-\eta}{\lambda +\eta} \right) + \right.  \nonumber\\
&&\left. + 2(\lambda -\eta) \int_0^1 \frac{\ln (1- t)\,  dt}{[2\eta
+ (\lambda -\eta)t ]^3} + \cal{O}(\varepsilon) \right\} \,   .
\label{eq25}
\end{eqnarray}

The analogous matrix element involving the reduced Green's function
can be evaluated by making use of the definition \eqref{eq16}:
\begin{equation}
\langle \bm{q} |G_{\mathrm{R}}(E_{1s}) V_{i\lambda}| 0 \rangle =
\lim_{\delta \to 0} \left\{ \langle \bm{q} |G_{\mathrm{C}}(E_{1s} -
\delta) V_{i\lambda}| 0 \rangle  + \frac{1}{\delta} \langle \bm{q}|
\psi_{1s}  \rangle \langle \psi_{1s}| V_{i\lambda}|0  \rangle
\right\}  \,   . \label{eq26}
\end{equation}
The counter-term in Eq.~\eqref{eq26} is given by
\begin{equation}
\frac{1}{\delta} \langle \bm{q}| \psi_{1s}  \rangle \langle
\psi_{1s}| V_{i\lambda}|0  \rangle = \frac{2^5 \pi m \eta^2}{
\varepsilon  q^4 (\lambda +\eta)^2} \,  .   \label{eq27}
\end{equation}
Here the explicit expression for the matrix element
\begin{equation}
\langle \psi_{1s}| V_{i\lambda}|0  \rangle = N_{1s}
\frac{4\pi}{(\lambda +\eta)^2 }
\end{equation}
and Eq.~\eqref{eq10} have been employed.

Adding Eqs.~\eqref{eq25} and \eqref{eq27}, one observes that the
pole terms cancel each other. Ac\-cor\-ding\-ly, we arrive at the
following expression
\begin{eqnarray}
\langle \bm{q} |G_{\mathrm{R}}(E_{1s}) V_{i\lambda}| 0 \rangle &=& -
2^5 \pi m \frac{\eta^2}{q^4}\left\{\frac{3}{2(\lambda +\eta)^2} +
\right.   \nonumber\\
&& \left. + \frac{(\lambda -\eta)}{(\lambda +\eta)^3}  - 2 (\lambda
-\eta) \int_0^1 \frac{\ln (1- t) \, dt}{[2\eta + (\lambda -\eta)
t]^3} \right\} \,   .   \label{eq29}
\end{eqnarray}
Using Eq.~\eqref{eq29} allows one to evaluate analytically the
matrix element entering  Eq.~\eqref{eq18}. It yields
\begin{equation}
N_{1s}^2 \left( - \frac{\partial}{\partial \mu} \right)
\frac{1}{\mu^2} \langle \bm{q} |  G_{\mathrm{R}}(E_{1s}) \left(
V_{i\eta} - V_{i(\eta +\mu)} \right)| 0 \rangle_{\left| \mu = 2\eta
\right. } =  \frac{m}{q^4} a_1 \,  .
\end{equation}
Here the coefficient $a_1$ appears as
\begin{equation}
a_1= -\frac{19}{16} + \frac{3}{4}\ln 2 \simeq - 0.6676
 \,   . \label{eq31}
\end{equation}

The next-to-leading-order correction ${\cal A}^{(1)}$ to the
amplitude of the single photoionization of helium-like ion in the
ground state is given by
\begin{equation}
{\mathcal A}^{(1)} =  \sqrt{2}\, N_{1s} N_\gamma \frac{4\pi \eta}{Z
q^4} \left(\bm{\mathrm e} \cdot \bm{p} \right) a_1 \,   .
\end{equation}
Employing Eqs.~\eqref{eq9} and \eqref{eq10} yields the total
amplitude
\begin{equation}
{\mathcal A} = {\mathcal A}^{(0)} + {\mathcal A}^{(1)} = {\mathcal
A}^{(0)}\left( 1 + \frac{a_1}{2Z}\right) \,   ,
\end{equation}
which accounts for the electron correlations. Within the same
approximation, the total cross section for the single K-shell
photoionization reads
\begin{equation}
\sigma^+ = \sigma_0^+ \left( 1 + a_1 Z^{-1}\right) \,  ,
\label{eq34}
\end{equation}
where $\sigma^+_{0}$ is given by Eq.~\eqref{eq12}.

The negative sign of the coefficient \eqref{eq31} can be understood
on the qualitative ground. It is well known that the single
photoeffect does not proceed on the free electron, while it can
occur on the bound one \cite{7}. The explanation of this fact
follows from the relation \eqref{eq11}: the nucleus serves as an
absorber of the recoil momentum $\bm{q}$. In the asymptotic region,
the value of $\bm{q}$ is relatively large, since the condition $q
\gg \eta$ holds. The Coulomb calculation performed within the
approximation of non-interacting electrons overestimates the cross
section, because the electron-nucleus bindings are utmost strong in
this case. Accounting for the screening effects attenuates these
bindings, so that, the cross section of the single photoionization
reduces in the absolute value.

As we already  mentioned, if the photon energy is not too high, the
cross section $\sigma^+_{\mathrm{C}}$ is more preferable rather than
$\sigma^+_{0}$. Due to slow convergence of the expansion of
Eq.~\eqref{eq13} with respect to the parameter $\xi$, it is still
legitimate to use of the following formula
\begin{equation}
\sigma^+ = \sigma_{\mathrm{C}}^+ \left( 1 + a_1 Z^{-1}\right)
\label{eq35}
\end{equation}
instead of Eq.~\eqref{eq34}.

As a testing ground for our analytical results, we choose the
neutral He atom, which seems to be the most thoroughly investigated
two-electron system. However, although extensive numerical
calculations of the photoionization cross sections have been
published in the literature, there are significant disagreements
between predictions based on different sophisticated methods at high
photon energies. The difficulty of comparison with experimental data
arises due to presence of additional contributions from the
scattering channels. One measures the total attenuation cross
section, but not exclusively the photoionization one \cite{1,2}. As
the most accurate theoretical calculations of the photoionization
cross sections at high-energy domain, Samson {\it et. al.} \cite{1}
have selected results of the work \cite{13}. Bell and Kingston used
the Hartree-Fock wave function for the continuum state and
many-parametrical variational wave function for the ground state. At
$\omega \simeq  3.5$ keV their result for the cross section is equal
to $5.74$ b, which however exhibits a gauge dependence on the level
of about $6\%$ \cite{13}. Our analytical formulas \eqref{eq34} and
\eqref{eq35} yield $8.24$ b and $5.73$ b, respectively. In this
case, the parameter $\xi =1/8$, but the value $\pi \xi$ is not too
small.

In the double photoeffect, one is usually interested in the ratio of
double-to-single ionization cross sections. At high photon energies,
the calculations performed within the framework of the leading-order
perturbation theory yield
\begin{equation}
R_0 = \frac{\sigma^{++}_0}{\sigma^+_0} = \frac{B}{Z^2} \,  ,
\label{eq36}
\end{equation}
where $B=0.090$ \cite{14,15} and $\sigma^+_{0}$ is given by
Eq.~\eqref{eq12}. Taking into account the higher-order screening
corrections to the total cross sections leads to the following
expression for the universal asymptotic ratio
\begin{equation}
R= R_0 \frac{(1 + b_1 Z^{-1} + b_2 Z^{-2} + \ldots)} {(1 + a_1
Z^{-1} + a_2 Z^{-2} + \ldots)} \,  . \label{eq37}
\end{equation}
The factor $R_0$ is separated out here, since the electron binding
energy is supposed to be negligibly small compared to the photon
energy. To any given order of the perturbation theory with respect
to the electron-electron interaction the representation \eqref{eq37}
is the Pad\'{e} approximant. The next-to-leading-order coefficient
$a_1$ is given by Eq.~\eqref{eq31}, while the other coefficients
remain to be calculated. Nevertheless, employing experimental data
for the double-to-single photoionization ratio, one can deduce an
estimate for the value of the coefficient $b_1$. The latter can be
obtained by equating the experimental value $R^{\mathrm{exp}} =
1.72(12)\%$ measured for He atom \cite{3} and the theoretical ratio
\eqref{eq37} truncated with taking into account only the
next-to-leading-order correlation corrections. It yields
\begin{equation}
b_1= - 0.981(71)  \,  .   \label{eq38}
\end{equation}

Having fixed the coefficients $a_1$ and $b_1$, we have calculated
the double-to-single photoionization ratio for helium isoelectronic
sequence. In Table \ref{table1}, we present a comparison of our
next-to-leading-order predictions according to Eq.~\eqref{eq37} with
the numerical results obtained by Forrey {\it et al.} \cite{4}. The
account of the screening corrections improves significantly the
asymptotic behavior for the leading-order ratio $R_0$ in the case of
light two-electron systems. This supports the statement concerning
the fast convergence of the $1/Z$ expansion even in the extreme
nonrelativistic domain \cite{16}. Indeed, the starting approximation
of the perturbation theory (assuming non-interacting electrons)
would be utmost inadequate for the description of light helium-like
ions, which are highly correlated. The relatively large values for
the coefficients \eqref{eq31} and \eqref{eq38} allow to correct the
zeroth approximation. Note also that the double-to-single
photoionization ratio \eqref{eq37} turns out to be less sensitive to
the higher-order screening corrections rather than the total cross
sections, since the coefficients $a_1$ and $b_1$ have the same sign.
Unfortunately, the significant uncertainty of the coefficient
\eqref{eq38} distorts the true behavior of the ratio $R$ in the case
of H${}^-$ ion. The coefficients $a_2$, $b_1$, and $b_2$ should be
calculated exactly within the framework of the consistent
perturbation theory.

Nevertheless, it is worthwhile to trace out the nontrivial behavior
of the series over the parameter $1/Z$ taking the binding energy for
the ground state in H${}^-$ ion as an example. Without the
electron-electron interaction the Coulomb binding energy is equal to
$27.2116$ eV. The corrections due to one-, two-, and three-photon
exchange diagrams are known to yield $-17.0079$ eV, $4.2921$ eV
\cite{17,18}, and $-0.1723$ eV \cite{18}, respectively. Then the
total binding energy turns out to be equal to $14.324$ eV. This
should be compared with the exact numerical result of $14.361$ eV,
which has been obtained within the approximation of an infinitely
heavy nucleus \cite{19}. On the level of accuracy of about $10^{-2}$
eV one already needs to take into account the effect of  nuclear
recoil. As seen, the terms of the $1/Z$ expansion exhibit
sign-changing oscillations and decrease fast in their absolute
value. Although the formal parameter of the perturbation theory is
equal to 1, the actual expansion turns out to converge by one order
of magnitude due to hidden parameters of the theory.

Concluding, we have evaluated the single K-shell photoionization
cross section with taking into account the next-to-leading-order
correlation correction. This allows one to improve the accuracy of
analytical predictions for light helium-like ions at high-energy
domain. We have discussed the universal behavior of the
double-to-single photoionization ratio as well as the fast
convergence of the perturbation theory with respect the parameter
$1/Z$.

\acknowledgments

A.M. is grateful to the Dresden University of Technology for the
hospitality and for financial support from Max Planck Institute for
the Physics of Complex Systems. A.N. and G.P. acknowledge financial
support from DFG, BMBF, and GSI. This research was also supported in
part by RFBR (Grant no. 05-02-16914) and INTAS (Grant no.
03-54-3604).

\newpage

\begin{table}[hbt]
\caption{ \label{table1}  
The asymptotic ratios of double-to-single photoionization cross sections
(in $\%$) are tabulated for various nuclear charge numbers $Z$.  
The ratio $R_0$ is calculated using the leading-order perturbation theory 
\protect{\cite{14,15}}. The ratio $R$ is calculated according to the 
truncated expression \protect{\eqref{eq37}}, taking into account the
next-to-leading-order corrections with the coefficients
\protect{\eqref{eq31}} and \protect{\eqref{eq38}}. The numerical
calculations by Forrey {\it et. al.} \protect{\cite{4}} have been
performed with the use of the fully correlated variational wave
functions.}
\begin{center}
\vspace{0.5 cm}
\begin{tabular}{l | r@{}l | r@{}l  | r@{}l
| r@{}l | r@{}l | r@{}l |  r@{}l |  r@{}l |  r@{}l |  r@{}l} \hline
\multicolumn{1}{c|}{$Z$} & \multicolumn{2}{c}{$1$} &
\multicolumn{2}{|c|}{$2$} & \multicolumn{2}{c}{$3$} &
\multicolumn{2}{|c|}{$4$} &\multicolumn{2}{c}{$5$} &
\multicolumn{2}{|c|}{$6$} &\multicolumn{2}{c}{$7$} &
\multicolumn{2}{|c|}{$8$} &\multicolumn{2}{c}{$9$} &
\multicolumn{2}{|c}{$10$}  \\  \hline \multicolumn{1}{l|}{$R_0$,
Eq.~\protect{\eqref{eq36}}} & 9  & .00 & 2 & .25 & 1 &  .00 & 0  &
.56 & 0  & .36 & 0  & .25 &  0 &  .18 & 0 &  .14 & 0 & .11 &
0  &  .09  \\
\hline \multicolumn{1}{l|}{$R$, Eq.~\protect{\eqref{eq37}}} & 0 &
.50 & 1 &  .72 & 0 & .865 & 0 & .509 & 0  &  .334 & 0  &  .235 & 0 &
.175 & 0 & .135 & 0 & .107 &
0 &  .087  \\
\hline \multicolumn{1}{l|}{$R$, Ref.~\cite{4}} & 1 & .602 & 1 & .644
& 0  &  .856 &  0  &  .508 & 0  &  .334 & 0  &  .236 & 0 & .175 & 0
&  .135 & 0  &  .107 &
0 &  .087  \\
\hline
\end{tabular}
\end{center}
\end{table}

\end{document}